\documentclass{elsarticle}

\usepackage{siunitx}
\journal{Journal of \LaTeX\ Templates}

\begin{document}

\begin{frontmatter}

\title{Tessellated granular metamaterials with tunable elastic moduli}


\author[mymainaddress]{Nidhi Pashine}

\author[mymainaddress]{Dong Wang}

\author[mymainaddress]{Jerry Zhang}

\author[mymainaddress]{Sree Kalyan Patiballa}

\author[mymainaddress,mythirdaddress]{Sven Witthaus}

\author[mysixthaddress]{Mark D. Shattuck}

\author[mymainaddress,mythirdaddress,myfourthaddress,myfifthaddress]{Corey S. O'Hern}

\author[mymainaddress]{Rebecca Kramer-Bottiglio\corref{mycorrespondingauthor}}

\cortext[mycorrespondingauthor]{Corresponding author}


\address[mymainaddress]{Department of Mechanical Engineering \& Materials Science, Yale University, New Haven, Connecticut 06520, USA}
\address[mythirdaddress]{Department of Physics, Yale University, New Haven, Connecticut 06520, USA}
\address[myfourthaddress]{Department of Applied Physics, Yale University, New Haven, Connecticut 06520, USA}
\address[myfifthaddress]{Graduate Program in Computational Biology \& Bioinformatics, Yale University, New Haven, Connecticut 06520, USA}
\address[mysixthaddress]{Benjamin Levich Institute and Physics Department, The City College of the City University of New York,
New York, New York 10031, USA}

\begin{abstract}
Most granular packings possess shear moduli ($G$) that increase with the applied external pressure, and bulk moduli ($B$) that increase or remain constant with pressure. This paper presents ``tessellated'' granular metamaterials for which both $G$ and the ratio $G/B$ decrease with increasing pressure. The granular metamaterials are made from flexible tessellations forming a ring of closed cells, each containing a small number of solid particles. For under-constrained tessellations, the dominant contributions to $G$ and $B$ are the particle-particle and particle-cell interactions. With specific particle configurations in the cells, we limit the number of possible particle rearrangements to achieve decreasing $G$ as we increase the pressure difference between the inside and outside of the tessellation, leading to $G/B \ll 1$ at large pressures. We further study tessellated granular metamaterials with cells containing a single particle and many particles to determine the variables that control the mechanical response of particle-filled tessellations as a function of pressure. 
\end{abstract}


\end{frontmatter}

\section{Introduction}


Granular materials consist of collections of macroscopic particles that interact with each other through contact forces, resulting in bulk mechanical behaviors that range from fluid-like to solid-like~\cite{behringer_physics_2018,puckett_equilibrating_2013}. At low densities, granular particles can flow past each other in a fluid-like state; at high densities, they jam into a solid-like state. Even in their jammed state, particles in granular media can rearrange by rolling and sliding past one another, thus changing the contact network and the corresponding mechanical response of the system. The mechanical properties of jammed packings, such as the bulk $(B)$ and shear $(G)$ moduli, are dependent on these interparticle interactions and scale as power laws with the packing fraction of the system~\cite{ohern_jamming_2003}.

The ratio $G/B$ of the shear and bulk moduli quantifies the solidity of granular materials. A liquid has $G/B=0$ whereas solids have $G/B > 0$. Moreover, $G$ and $B$ are within an order of magnitude of each other for most materials~\cite{ashcroft_solid_1976, greaves_poissons_2011}. 
For example, the shear  and bulk moduli of iron are $77.6~\si{\giga\pascal}$ and $166~\si{\giga\pascal}$~\cite{noauthor_online_nodate}, respectively, with a $G/B \approx 0.47$. 
Further, for most metals~\cite{bridgman_compressibility_1923,peng_pressure_2005, brown_experimental_2020}, as well as ionic~\cite{cowley_theoretical_1990,kwon_pressure_1995} and non-ionic~\cite{ludwig_c60_1994} crystalline solids, $G$ and $B$ increase as a function of pressure. Most metals have a relatively constant value of $G/B$ over a wide range of pressures~\cite{partom_change_2017}. 
In contrast, granular packings with large numbers of particles show an increasing $G/B$ with increasing pressure above the jamming transition, due to particle rearrangements within the packing~\cite{ohern_jamming_2003,goodrich_solids_2014,still_phonon_2014}. 

This paper presents ``tessellated'' granular metamaterials with tunable elastic moduli and $G/B$ values dependent on pressure. The tessellated granular metamaterials are made of an annulus that is radially tessellated into cells, each of which is filled with particles in identical configurations that are close to the onset of jamming. 
Notably, by filling the tessellated cells with a small number of particles, we limit the number of possible rearrangements and thus increase our control over $B$, $G$, and $G/B$. 
Our results show that the $B$ of our tessellated granular metamaterial does not always increase monotonically with pressure, although does generally possess a positive correlation with pressure.
Due to the small number of possible local rearrangements in the cells, we also observe a global decrease in $G$ with increasing pressure. Combined, we find that $G/B$ for the tessellated granular metamaterials decreases with increasing external pressure.

\begin{figure}
    \centering
    \includegraphics[width = 90mm]{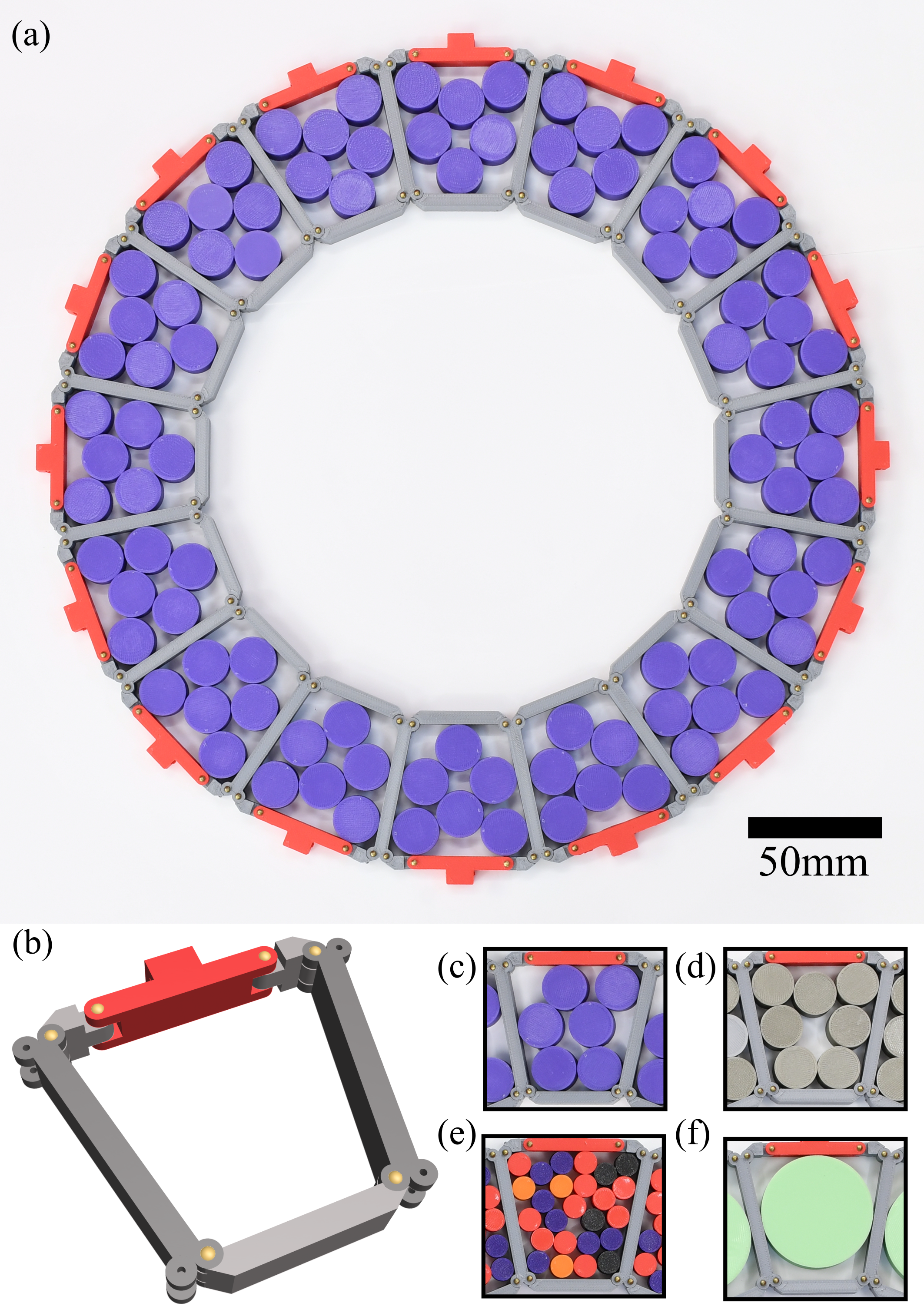}
    \caption{(a) Experimental setup. 3D printed tessellation with flexible joints filled with particles. The particles as well as the tesselation are 3D printed out of Polylactic Acid (PLA). Particles in each cell have the same configuration, referred to as configuration `6 particles I'. (b) A single cell of the tessellation showing the edges and joints. The different particle configurations studied in this work include: (c) A single cell with particles in configuration `6 particles II,' (d) 7 particle configuration, (e) 25 particles per cell with each cell in a random particle configuration, and (f) one particle in each cell with varying particle diameter.}
    \label{fig1:design}
\end{figure}

\section{Methods}

\subsection*{Tessellation Geometry}
Our setup includes a flexible tessellation with walls that can rotate freely about its joints to create a system with a large number of zero-energy modes. According to Maxwell's counting argument, a 2D tessellation with $N$ nodes and $N_b$ bonds has $(2 N - N_b)$ zero energy modes, including three trivial modes corresponding to translation and rotation~\cite{thorpe_continuous_1983,maxwell_l_1864}. As shown in Fig.~\ref{fig1:design}(a), our tessellation design consists of 16 trapezoid cells that are connected together to form an annulus. Fig.~\ref{fig1:design}(b) shows the structure of a single cell. Each cell has four joints on the vertices of the trapezoid, and two additional joints breaking up the top wall into three parts. These extra joints add two more degrees of freedom per cell to the system. When connected together in an annulus, the tessellation has $N$~=~64 nodes and $N_b$~=~80 bonds, resulting in 45 zero-energy modes in the system. When built in an experiment, this structure is highly flexible, can undergo large deformations, and is fully collapsible. We fill each of the cells with particles in configurations that are derived via discrete element method (DEM) simulations.

\begin{figure}
    \centering
    \includegraphics[width = 120mm]{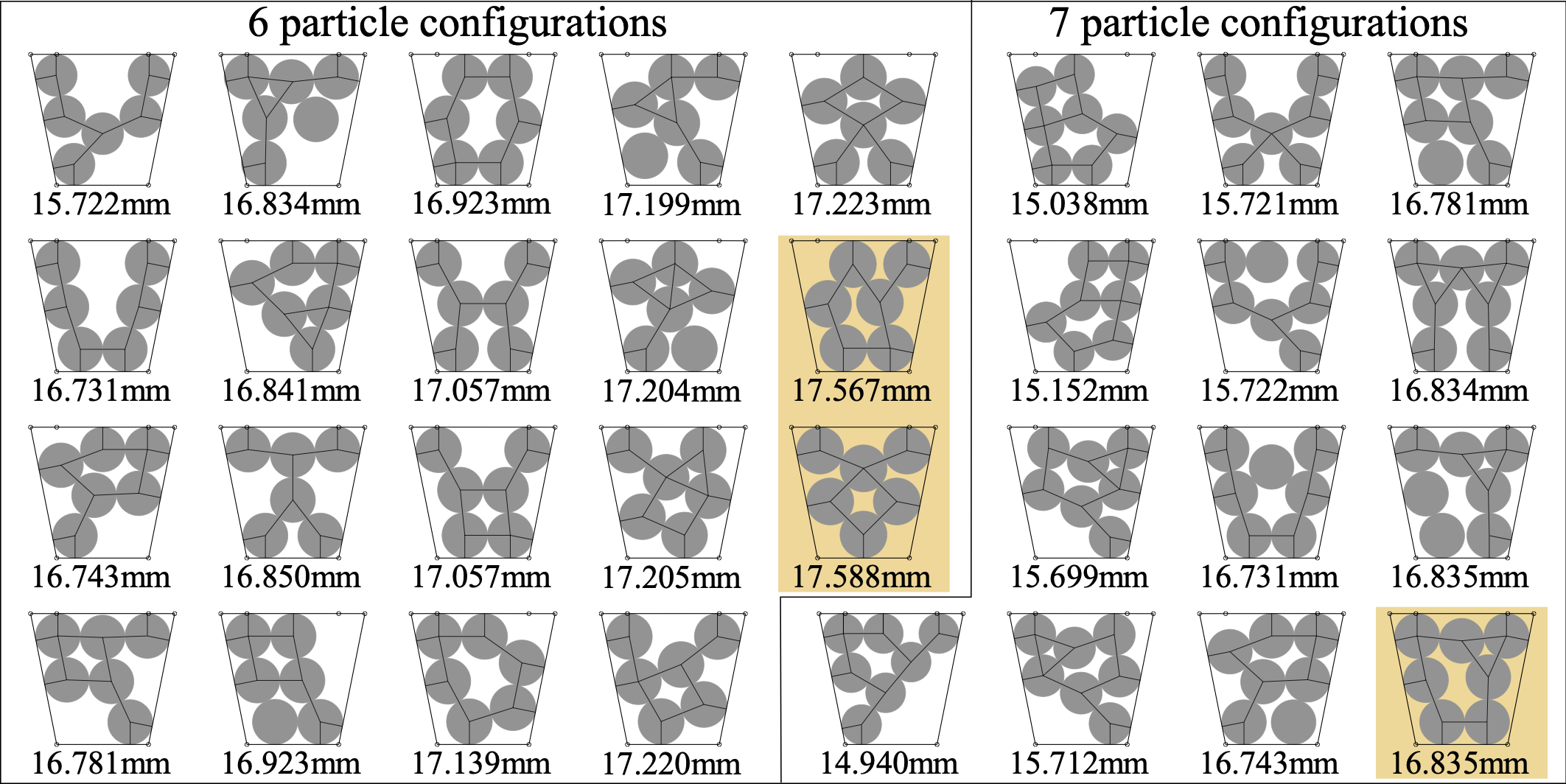}
    \caption{All unique $6$ and $7$ particle-filled cells obtained from the DEM simulations with the particle sizes for each configuration. Lines originating from particle centers correspond to particle-particle or particle-wall contacts. The configurations studied in the experiments correspond to the largest packing fraction and are highlighted in yellow.}
    \label{fig:simulations}
\end{figure}

\subsection*{Discrete Element Method Simulations}
To enumerate all possible jammed packings within a cell of the tessellation, we employ discrete element method (DEM) simulations for $N$ monodisperse, frictionless disks in the geometry shown in Fig.~\ref{fig1:design}(b). Endpoints of a cell are held fixed while generating a jammed disk packing. Disks interact with each other via the pairwise, purely repulsive linear spring potential energy:
\begin{equation}
    \label{eq:energy_pp}
    U^{pp}_{jk} = \frac{\epsilon_{pp}}{2} \left(1 - \frac{r^{pp}_{jk}}{\sigma_{jk}} \right)^2 \Theta \left(1 - \frac{r^{pp}_{jk}}{\sigma_{jk}}\right),
\end{equation}
where $\epsilon_{pp}$ is the characteristic energy scale of the repulsive interactions, $r^{pp}_{jk}$ is the distance between the centers of disks $j$ and $k$, $\sigma_{jk}$ is the sum of the radii of disks $j$ and $k$, and $\Theta(\cdot)$ is the Heaviside step function. The interaction between the $i$th side and the $j$th disk is also purely repulsive and given by:
\begin{equation}
    \label{eq:energy_pb}
    U^{pb}_{ji} = \frac{\epsilon_{pb}}{2} \left(1 - \frac{r^{pb}_{ji}}{R_j} \right)^2 \Theta \left(1 - \frac{r^{pb}_{ji}}{R_j}\right),
\end{equation}
where $\epsilon_{pb}$ is the particle-wall stiffness, $r^{pb}_{ji}$ is the shortest distance between the center of disk $j$ and the $i$th side, and $R_j$ is the radius of disk $j$.

We measure the stress tensor $\Sigma_{\alpha \beta}$ in the system via the Virial expression. $\Sigma_{\alpha \beta} = \Sigma^{pp}_{\alpha \beta} + \Sigma^{pb}_{\alpha \beta}$ includes two terms representing the particle-particle interactions, $\Sigma^{pp}_{\alpha \beta}$; and particle-boundary interactions, $\Sigma^{pb}_{\alpha \beta}$. 

The particle-particle stress $\Sigma^{pp}_{\alpha \beta}$ is given by
\begin{equation}
    \label{eq:stress_pp}
    \Sigma^{pp}_{\alpha \beta} = \frac{1}{A}\sum^N_{j,k}f^{pp}_{jk\alpha}r^{pp}_{jk\beta},
\end{equation}
where $A$ is the area of the confining boundary, $f^{pp}_{jk\alpha}$ is the $\alpha$ component of the force on disk $j$ from disk $k$, and $r^{pp}_{jk\beta}$ is the $\beta$ component of the separation vector from the center of disk $k$ to the center of disk $j$. 

The particle-boundary stress $\Sigma^{pb}_{\alpha \beta}$ is given by
\begin{equation}
    \label{eq:stress_pb}
    \Sigma^{pb}_{\alpha \beta} = \frac{1}{A} \sum^4_i \sum^N_{j}f^{pb}_{ji\alpha}r^{pb}_{ji\beta},
\end{equation}
where $f^{pb}_{ji\alpha}$ is the $\alpha$ component of the force on disk $j$ from side $i$ of the boundary, and $r^{pb}_{ji\beta}$ is the $\beta$ component of the separation vector from the contact point between side $i$ and disk $j$ to the center of disk $j$. From the stress tensor $\Sigma_{\alpha\beta}$, we obtain the pressure $P = (\Sigma_{xx} + \Sigma_{yy}) / 2$, the shear stress $\Sigma = -\Sigma_{xy}$ in simple shear, and $\Sigma = (\Sigma_{xx} - \Sigma_{yy}) / 2$ in pure shear.

To generate a jammed disk packing in the tessellation, we start with a dilute system where disks are randomly placed in one cell and replicated in the rest of the cells, with the packing fraction at $\phi < 10^{-3}$. We then increase the particle sizes by $\Delta \phi/\phi = 2 \Delta R/R = 2 \times 10^{-3}$, followed by energy minimization using the fast inertia relaxation engine (FIRE) algorithm \cite{bitzek_structural_2006}. After energy minimization, we measure the internal pressure $P = (\Sigma^{pp}_{xx} + \Sigma^{pp}_{yy} + \Sigma^{pb}_{xx} + \Sigma^{pb}_{yy}) / 2$ for any given cell. Since all cells start with the same initial configuration, the pressure, $P$ in any cell is the same. If $P$ is smaller than the target pressure, $P_t = 10^{-7}$, we grow the particles by $\Delta \phi$ again and apply energy minimization. If $P > P_t$, we then return to the disk and boundary configuration before the last growth step and increase $\phi$ by $\Delta \phi/2$. We repeat this search procedure until we reach a state with $|P- P_t| / P_t < 10^{-4}$. We generate $10^5$ jammed packings for each $N$ with different random initial configurations and find all the unique particle-filled cells. The unique particle configurations for $6$ and $7$ particle systems obtained from the simulations are shown in Fig.~\ref{fig:simulations}.

\subsection*{Experimental design}

The experimental system is created by 3D printing the tessellation and particles. From the simulated jammed configurations, we choose the particle arrangements with the largest packing fractions for $6$ and $7$ particles per cell, which are highlighted in Fig.~\ref{fig:simulations}. Even when we directly translate these simulated configurations to experiments, there are several differences between the simulations and the experimentally recreated packings. Instead of growing particles inside a cell, we place them by hand, leading to differences in particle positions and geometry of the cells. Additionally, due to the presence of a large number of zero-energy modes, each cell can have a slightly different geometry. These slight deviations from the designed configurations are enough to un-jam the whole system. When we apply compressive or shear deformations to the system, we notice that the initial response at small strains is dominated by frictional interactions, but as we increase the strain, the system reaches a jammed state. 

The different particle configurations we study are shown in Fig.~\ref{fig1:design}. We build three designed configurations with $6$ and $7$ particles in each cell. Additionally, we also compare these results to a system with a large number of particles ($25$ particles) in random configurations per cell, and systems with a single particle per cell. For the single-particle systems, we study the effect of the variation of particle size on the mechanical response of the tessellation.
Fig.~\ref{fig1:design}(a) shows the tessellation filled with a configuration of $7$ particles in each cell. Our experimental design is 2D, and both the tessellation walls and the disk-shaped particles have a height of $10\si{\milli\meter}$ in the third dimension. The tessellation and particles are made of Polylactic Acid (PLA), 3D printed using a Prusa i3 MK3S printer and a print infill of $20\%$. The experiments are conducted with the tessellation on a horizontal surface of Polyethylene terephthalate (PET). The PET sheet is covered with a thin layer of corn starch to minimize any frictional effects between the particles and the substrate. The magnitude of frictional forces between the experimental system and the substrate is estimated by measuring the minimum force required to translate a filled tessellation across the substrate and is $\mathcal{O}(1\si{\newton})$. Details on the dimensions and geometry of the walls, as well as the particles, are included in the SI.

\section{Results}
\subsection*{Bulk modulus}
\begin{figure}
    \centering
    \includegraphics[width = 120mm]{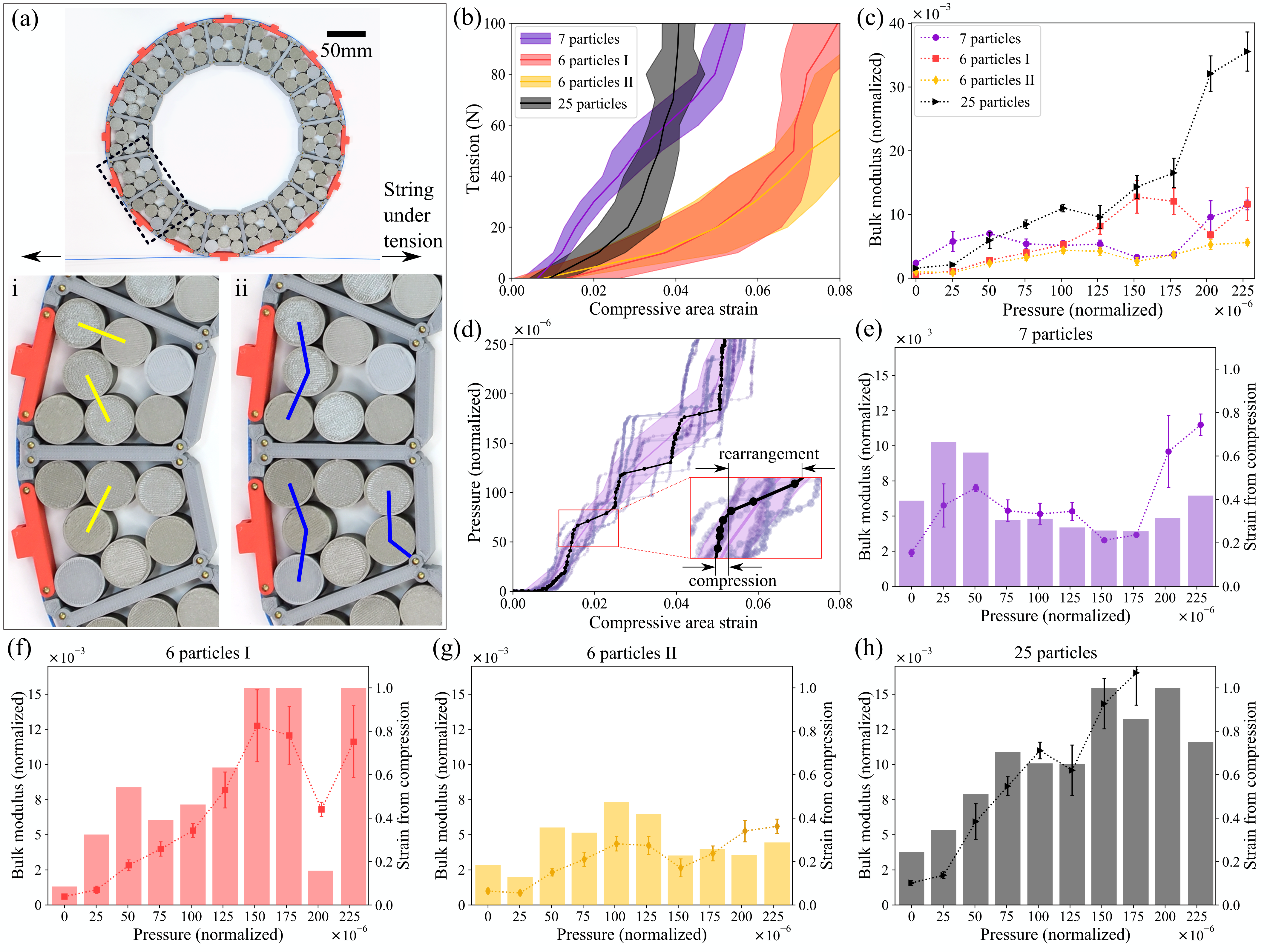}
    \caption{(a) Bulk compression experiments in the configuration with 7 particles per cell. A cord going around the tessellation is pulled under tension and the resulting change in area gives the bulk modulus of the system. When under pressure, particles and cells rearrange to create different contact networks. (i) System before a rearrangement event---yellow lines show contacts that disappear after rearrangement. (ii) System after a rearrangement---blue lines show new contacts formed after the rearrangement. (b) Cord tension vs. compressive area strain of the annulus. The slopes of the curves give the bulk modulus. (c) Bulk modulus at different external pressures. (d) Individual measurements and average for the $7$ particle configuration. Data highlighted in black to show two different behaviors: particle compression (large slope) and rearrangements (small slope). (e) Measure of how much strain change in the $7$ particle system occurs from rearrangements. Bar chart derived from plot (d) shows the average fraction of strain change that occurs due to particle compression at different pressures. Bar chart is overlayed by the bulk modulus as a function of pressure. (f-h) Same plot for the 6 and 25 particle configurations.}
    \label{fig2:bulk}
\end{figure}

In the 2D tessellation, the bulk modulus is defined as $B = -\frac{dP}{d\gamma_{area}}$, where $P$ is the external pressure and $\gamma_{area}$ is the area strain. $\gamma_{area} = \frac{A^0-A'}{A^0}$, where $A^0$ is the initial area of the inner cavity of the tessellation and $A'$ is the internal area after applying external pressure, $P$. The experimental setup to measure the bulk modulus is shown in Fig.~\ref{fig2:bulk}(a). To simulate external pressure, we run a cord around the outer edge of the tessellation and apply a known tension to the cord. This cord tension $(T)$ exerts a normal force on the outer edge of the tessellation, which acts as a simulated external pressure $(P)$. The relation between $T$ and $P$ is derived in the SI and is given by $P = \frac{Tf}{Rh}$, where $f$ is the fraction of the outer circumference that is in contact with the cord, $R$ is the outer radius, and $h$ is the height of the tessellation in the third dimension. 

The system responds to external pressure by decreasing its internal area. As the external pressure increases, we see intermittent particle rearrangements in the tessellation cells. One such rearrangement is shown in Fig.~\ref{fig2:bulk}(a)i-ii. During a rearrangement event, certain particle contacts that existed in a configuration are broken and new contacts are formed. These rearrangements, which change particle-particle as well as particle-wall contacts, often modify the shape of individual cells, which subsequently changes the internal area of the tessellation. 

To measure the pressure response of the system, we quasi-statically increase the applied tension in the cord and measure the change in the area of the tessellation. Fig.~\ref{fig2:bulk}(b) shows the cord tension in the system as a function of compressive strain. As expected, higher tension in the cord leads to higher compressive strains. The maximum compressive strain that the system can undergo depends on the packing fraction $(\phi)$ of the configuration. Among the configurations studied, $\phi$ increases slightly with increasing numbers of particles in the cells $(\phi_{6p} = 0.699, \phi_{7p} = 0.747, \phi_{25p} = 0.763)$.  Since the tension in the cord is proportional to external `pressure,' the local slope of this curve at any given value of pressure (tension) is a measure of the bulk modulus of the system at that pressure. Fig.~\ref{fig2:bulk}(c) shows bulk modulus as a function of external pressure. Both pressure and bulk modulus are normalized by the average Young's modulus of the particles. Details on the estimation of Young's modulus are in the SI. 

As seen in Fig.~\ref{fig2:bulk}(c), the bulk modulus of each particle configuration is different and the bulk modulus can change non-monotonically with pressure, although it mostly increases with increasing pressure.
To better understand the relation between bulk modulus and particle rearrangements, we study the 7-particle configuration as a sample system. Fig.~\ref{fig2:bulk}(d) overlays the averaged data (shaded region) and individual trials (curves with points) of pressure vs. compressive area strain. One randomly chosen curve is highlighted in black. At very small strains ($\gamma_{area}<1\%$), almost no force is required to strain the system because the particles are unjammed. At the onset of jamming, each experiment presents in a staircase-like pattern, which captures compression regimes where the particles compress against each other (vertical lines) and undergo rearrangement events (horizontal lines). Studying each compression and rearrangement event (Fig.~\ref{fig2:bulk}(d) inset), we can quantify the contributions to the strain change as coming from compression $(\gamma_C)$ or rearrangements $(\gamma_R)$. Fig.~\ref{fig2:bulk}(e) shows the strain from compression, $\frac{<\gamma_C>}{<\gamma_C>+<\gamma_R>}$, as a function of pressure in the form of a bar chart, where $<\gamma_C>$ and $<\gamma_R>$ are averages over 10 experiments. Fig.~\ref{fig2:bulk}(e) also shows the bulk modulus at corresponding pressures, which is highly correlated with the rearrangement data. This correlation persists in all the multi-particle configurations we studied, as shown in Fig.~\ref{fig2:bulk}(f)-(h), implying that particle rearrangements decrease $B$ locally. However, rearrangements often lead to a more stable packing, which leads to an overall increase in $B$ with pressure.
At higher pressures, the correlation between rearrangements and bulk modulus goes down, as seen somewhat in the 7-particle case (Fig.~\ref{fig2:bulk}(c)) but more acutely in the 25-particle case (Fig.~\ref{fig2:bulk}(h)). 
We suspect that at very high pressures, there are fewer rearrangements possible and that packings with more particles find increasingly stable configurations that contribute to $B$.

\subsection*{Shear modulus}

\begin{figure}
    \centering
    \includegraphics[width = 120mm]{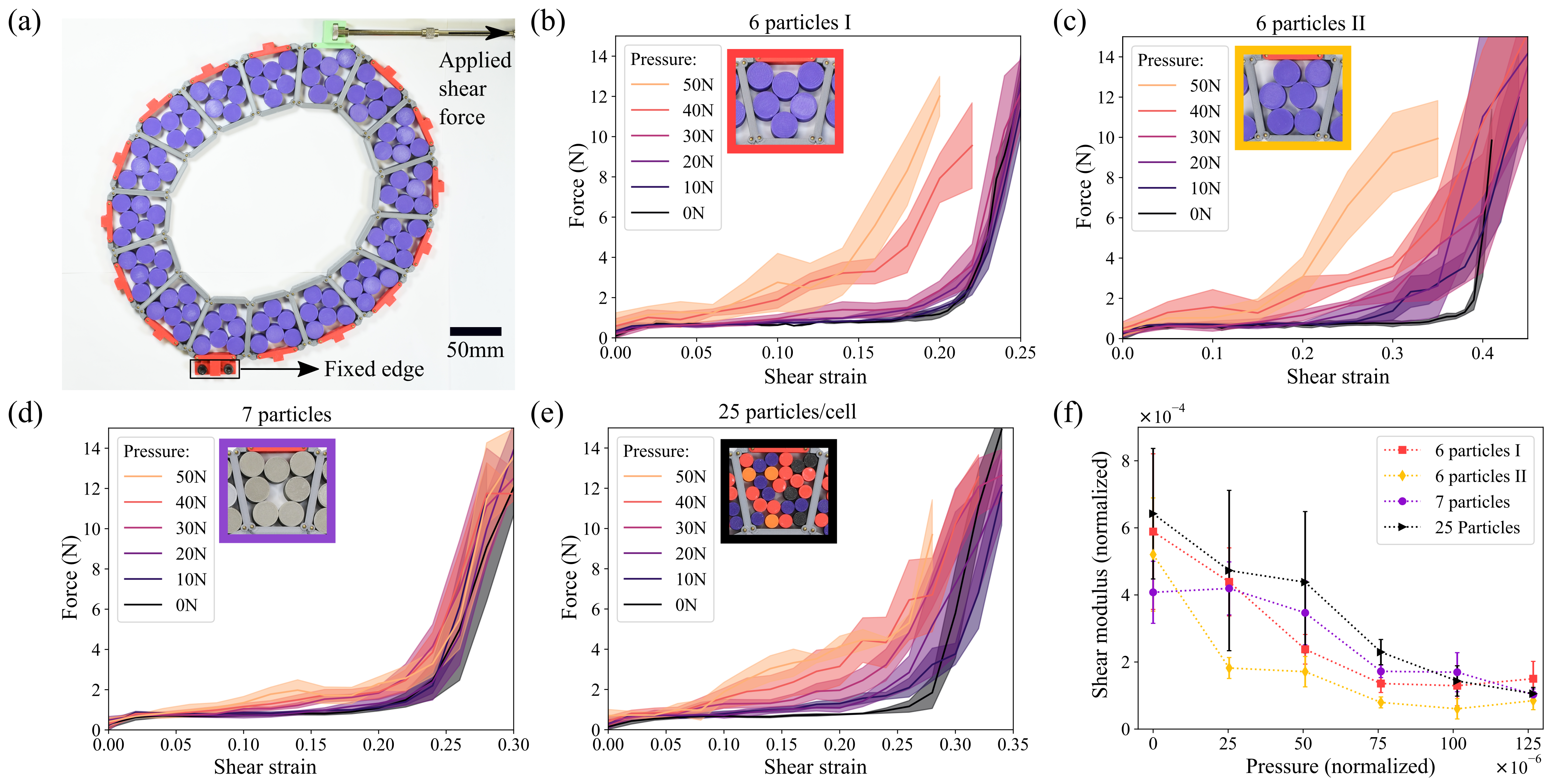}
    \caption{Force response to applied shear strain at different external pressures. (a) 7-particle system, (b) 25-particle system with random configurations in each cell, (c) 6-particle configuration I, (d) 6-particle configuration II, and (e) 25-particles per cell in random configurations. (f) Shear modulus at different external pressures.}
    \label{fig3:shear}
\end{figure}

To measure the shear modulus of our tessellated granular metamaterial, we fix the straight outer wall element of the bottom cell and apply a tangential force on the straight outer wall of the top cell, as shown in Fig.~\ref{fig3:shear}(a). In a simple shear measurement of a continuum material, the horizontal displacement of the top wall is proportional to the applied simple shear strain. In our system, simple shear strain is difficult to measure because a tangential force on the top wall rotates and deforms different parts of the annulus in different ways. Therefore, we fit an ellipse to the inner cavity of the annulus and use its dimensions to calculate the shear strain applied to the system. 
The shear strain in the system is given by $\gamma_{shear} =\frac{a-b}{\sqrt{2}r}$, where $a$ and $b$ are the major and minor axes of the ellipse fitted to the inner cavity and $r$ is the radius of the initial circular cavity. The shear stress is given by $\sigma = \frac{F}{2R\times h}$, where $F$ is the applied shear force and $2R\times h$ is the cross-sectional area of the system. Finally, we arrive at the shear modulus, which is given by $G = \frac{d\sigma}{d\gamma_{shear}}$. Just as for the bulk modulus, $B$, the reported values of $G$ are normalized by Young’s modulus of the particles. The relations for shear strain, shear stress, and shear modulus are derived in the SI.

The shear response of our system has two distinct regimes, as shown in Fig.~\ref{fig3:shear}(b-e). At small shear strains, the system is unjammed and the response is dominated by particle-particle and particle-cell frictional interactions. Upon shear-jamming, the tessellation requires substantially larger forces to further shear.
As expected, as the external pressure on the tessellation increases the tessellation shear jams at smaller strains.

To estimate the shear modulus of each configuration, we measure the slope of the curves in Fig.~\ref{fig3:shear}(b-e) in a strain regime where the system is shear jammed, at force values between $2\si{\newton}$ and $10\si{\newton}$. The lower force limit of $2\si{\newton}$ is to ensure that the shear modulus measurement is in the shear jammed regime and larger than the frictional forces between the tessellation and the substrate. Fig.~\ref{fig3:shear}(f) shows the shear modulus for each configuration as a function of pressure, averaged over five measurements. The shear modulus for all of the studied configurations decreases monotonically with pressure, which is uncharacteristic of granular materials. 

Prior work has shown that, for a given jammed packing, $G$ decreases with pressure until there is a change in the contact network~\cite{vanderwerf_pressure_2020, wang_shear_2021}. 
In our shear experiments, the $6$-particle configurations do not undergo particle rearrangements once the system has been shear jammed. The $7$- and $25$-particle systems show a few particle rearrangement events under high shear forces, but these minor rearrangements do not change the contact network substantially. Thus, all of the systems show a decrease in $G$ as a function of pressure. 
Note that the $25$-particle system is unlike the $6$- and $7$-particle designed systems, as the initial particle configurations are different in each cell and also different for each shear experiment. We correlate this more random nature of the $25$-particle system with the larger variation in $G$ (relative to the other systems with few particles per cell), especially at lower pressures. 

\subsection*{Single particle per cell}
\begin{figure}
    \centering
    \includegraphics[width = 120mm]{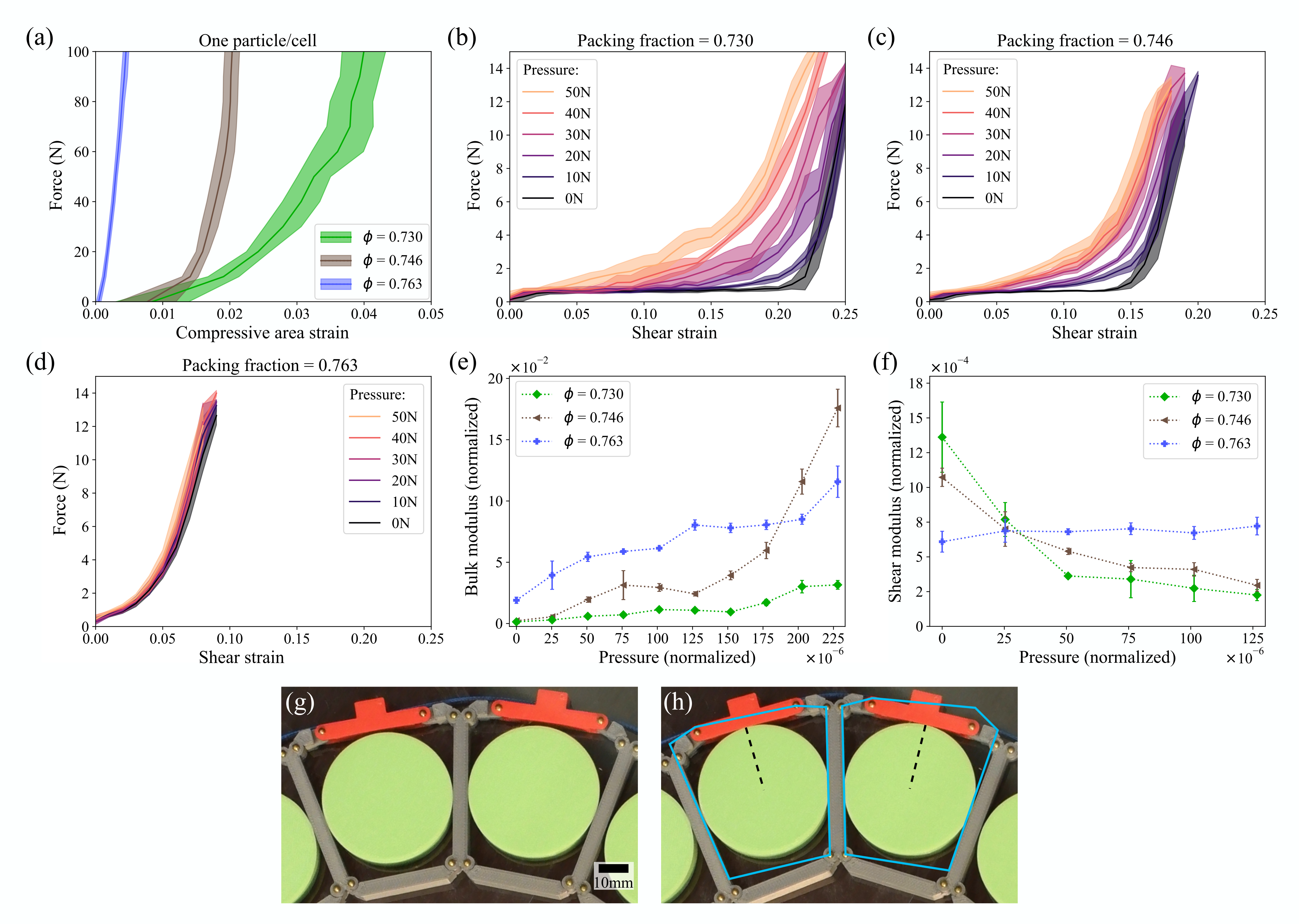}
    \caption{One particle system. (a) Force response to bulk compression for single particle systems with different packing fractions. Force vs shear strain at different external pressures for increasing packing fraction: (b) $\phi = 0.730$ (particle diameter = $44\si{\milli\meter}$) (c) $\phi = 0.746$ (particle diameter = $44.5\si{\milli\meter}$) (d) $\phi = 0.763$ (particle diameter = $45\si{\milli\meter}$). (d) Shear modulus at different external pressures. (e) Bulk modulus as a function of external pressure. (f) Shear modulus as a function of pressure. Rearrangement event in $\phi = 0.730$ system. Images show configurations (g) before a rearrangement (h) after a rearrangement. Dashed black line shows new contacts and solid blue line shows the initial geometry of the cells.}
    \label{fig4:single particle system}
\end{figure}

For a better understanding of our tessellated granular material, we investigate a simpler system with a single particle in each cell. This configuration has minimal direction dependence and the only rearrangements that take place involve the particle and tessellation walls. We study this one-particle-per-cell system by varying a single parameter, the particle size. By increasing the particle size, we increase the packing fraction in each cell.
As expected, we observe that varying the packing fraction substantially changes the bulk and shear moduli of the tessellation. By increasing the particle size and packing fraction, the overall flexibility of the tessellation decreases, which then limits the maximum compression or shear the system can undergo.

Nevertheless, even a single-particle-per-cell yields similar $B$ and $G$ trends to what we observed for the multi-particle systems. As shown in Fig.~\ref{fig4:single particle system}(a), as we increase the particle size the forces required to compress the system increase. Looking again for a relation between $B$ and particle rearrangements, we notice that the single-particle systems undergo rearrangements by changing the particle-wall contacts, as shown in Fig.~\ref{fig4:single particle system}(g-h). In the single-particle setup, as we increase the packing fraction, the initial system becomes more tightly packed, resulting in fewer possible rearrangements, and therefore a higher bulk modulus.
The bulk modulus measured from the slopes of the force vs strain curves in Fig.~\ref{fig4:single particle system}(a) are shown in Fig.~\ref{fig4:single particle system}(e). We see that the system with the largest packing fraction, $\phi = 0.763$, which does not undergo any rearrangements, has the largest bulk modulus. The system with intermediate packing fraction, $\phi = 0.746$, does undergo rearrangements, and the rate of increase of $B$ with pressure notably increases after the system has rearranged.

The shear response of single-particle systems (Fig.~\ref{fig4:single particle system}(b-d)) show that tessellations filled with larger particles shear jam at smaller strains. 
The effect of external pressure on the shear response also varies with particle size, as shown in Fig.~\ref{fig4:single particle system}(f).
For the smallest particle we tested ($\phi = 0.730$), the shear modulus decreases rapidly with increasing external pressure. 
For the largest particle we tested ($\phi = 0.763$), varying the external pressure does not change the shear response substantially. 
We surmise that, due to increased internal pressure (in the cells) from increased packing fraction, the material system is less susceptible to an externally applied pressure, and $G$ thus remains constant over a range of external pressures.

\subsection*{Ratio of Shear to Bulk Modulus $(G/B)$}

\begin{figure}
    \centering
    \includegraphics[width = 50mm]{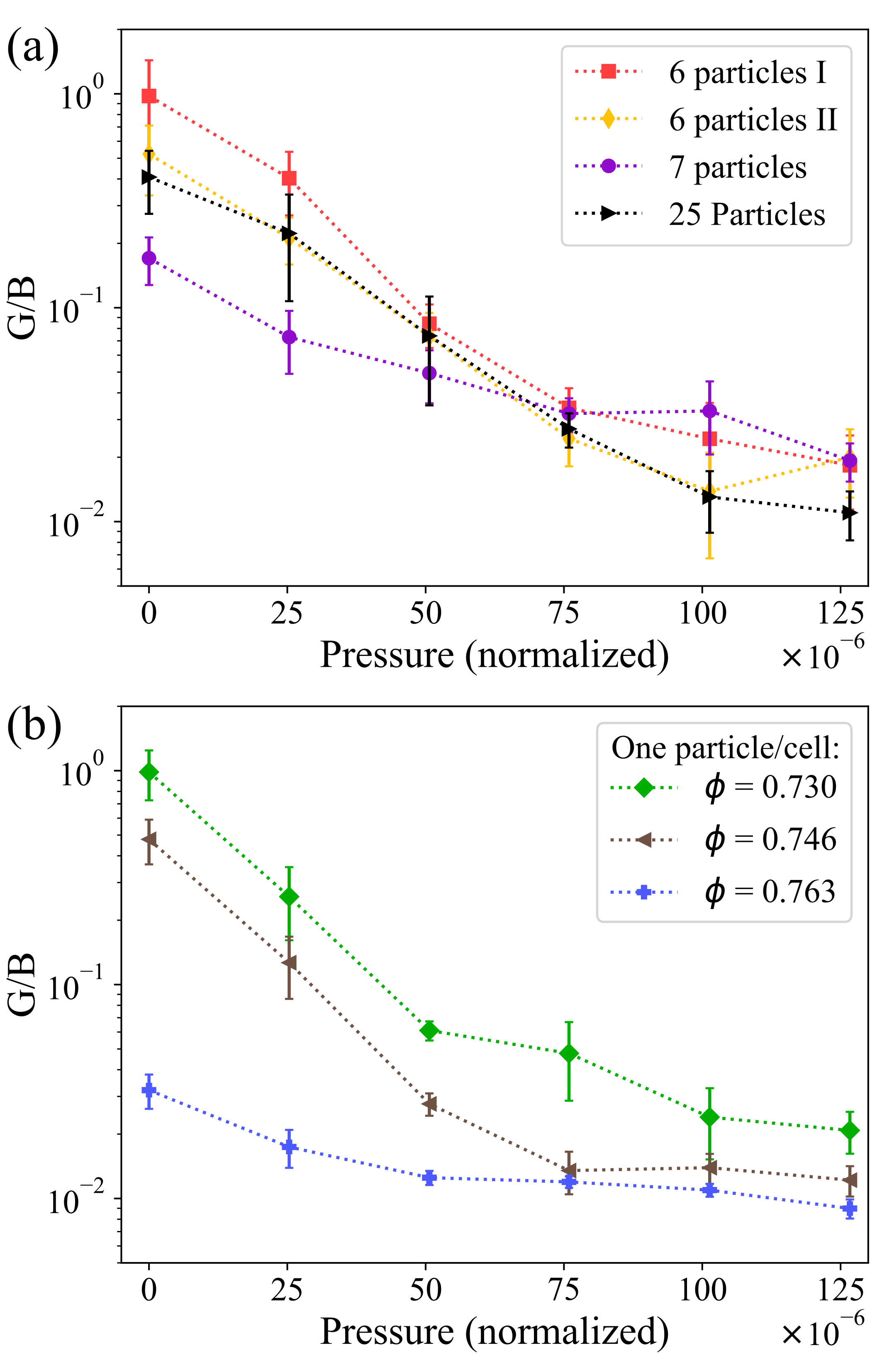}
    \caption{Ratio of shear to bulk modulus $(\frac{G}{B})$ for different particle configurations. (a) $\frac{G}{B}$ as a function of pressure for $4$ different particle configurations. All of the configurations show a monotonic decrease in the value of $\frac{G}{B}$ as a function of pressure. (b) $\frac{G}{B}$ as a function of pressure for a single particle per cell with increasing size of particles (\textit{i.e.}, packing fraction).} 
    \label{fig5:G_over_B}
\end{figure}

Based on the $B$ and $G$ measurements in the previous sections, we calculate the ratio of the two moduli, $G/B$, for different configurations. $G/B$ values for multi-particle systems are shown in Fig.~\ref{fig5:G_over_B}(a). All of the multi-particle configurations studied have $G/B = \mathcal{O}(1)$ at zero external pressure and the value of $G/B$ decreases consistently with increasing pressure.

The relation between $G/B$ and pressure in single particle systems, shown in Fig.~\ref{fig5:G_over_B}(b), varies with packing fraction. The response of low packing fraction systems $(\phi = 0.730$ and $\phi = 0.746)$ is similar to the multi-particle systems. In the previous section, Fig.~\ref{fig4:single particle system}(e) and (f) show that for single particle systems, $G$ decreases while $B$ increases as a function of pressure, leading to their ratio, $G/B$ decreasing with pressure. This response is different for the largest packing fraction, $\phi = 0.763$,
which starts in a jammed state and leaves no opportunity for particle rearrangements. 
Increasing the packing fraction of a given configuration effectively increases the internal pressure in each cell. Thus, even when no external pressure is applied, the system with $\phi = 0.763$ is under pressure and therefore has a small value of $G/B$.

\section*{Discussion}
Most prior studies of jammed granular packings have focused on systems with a large number of particles. In these systems, it is challenging to have the packing maintain the same interparticle contact network. As a result, studies with a large number of particles often focus on the ensemble averages of material properties. We show that by creating a tessellated granular system with a small number of particles in each cell, we can control the initial state, as well as limit the possible rearrangements in the system. Not only has this allowed us to create a material with a $G/B$ lower than most materials, but we have been able to reverse the typical pressure-dependent shear response in a granular material. 

We were motivated to design a material with a low $G/B$ for next-generation atmospheric diving suits (ADS). ADS have traditionally been designed as hard-shell submersibles, with complex anthropomorphic joints to allow articulation while maintaining an internal pressure of 1 atm. Despite significant progress in ADS, current designs are bulky and limit a diver’s mechanics of motion. Ambient pressure diving suits, such as scuba gear, allow for much greater diver maneuverability, but cannot protect the diver from extreme pressures and from the associated physiological problems. Therefore, new materials that are flexible but resistant to pressure gradients are needed to develop new ADS. Such material innovation would enable ADS that allow for free bending and twisting at body joints, while simultaneously protecting the diver from the hydrostatic pressures at large depths. This target application further motivated our annulus tessellation design, as it is a 2D representation of what can later be developed into a 3D wearable sleeve or joint. 

This work opens further questions regarding the design and performance of tessellated granular metamaterials. An exciting challenge is to design and build such a system in three dimensions. Our results suggest further studies to understand the limits of tessellated granular metamaterial tunability, cyclic stability, and reversibility. 
Herein, we focused on monodisperse packings of circular disks with repeated configurations in each tessellated cell, but the  behavior of granular systems depends on particle properties~\cite{brito_universality_2018,athanassiadis_particle_2013}. Future inquiries should include additional particle parameters such as particle shape, stiffness, and polydispersity, which would allow for further tuning of the mechanical response of tessellated granular metamaterials.

\section*{Acknowledgements}
This work was supported by the Office of Naval Research under Grant No. N00014-20-1-2640.

\end{document}